\documentstyle[prb,aps,12pt,preprint]{revtex}

\begin{document}
\draft
\title{Comparison of in- and out-of-plane charge dynamics in
YBa$_2$Cu$_3$O$_{6.95}$}
\author{E. Schachinger}
\address{Institut f\"ur Theoretische Physik, Technische Universit\"at
Graz\\A-8010 Graz, Austria}
\author{J.P. Carbotte}
\address{Department of Physics and Astronomy, McMaster University,\\
Hamilton, Ont. L8S 4M1, Canada}
\date{\today}
\maketitle
\begin{abstract}
The in-plane optical conductivity has been successfully employed
to obtain information about the coupling of the charge carriers
to the spin degrees of freedom in the high $T_c$ oxides. We
investigate how this inelastic scattering affects the out of
plane charge dynamics. We consider both coherent (in-plane
momentum is conserved) and incoherent (no momentum conservation)
$c$-axis transfers and find that the two cases give quite a
distinct $c$-axis conductivity as a function of energy $\omega$.
Comparison of our theoretical calculations with the available data
does not allow a definitive conclusion, but a momentum dependent
coherent matrix element characteristic of the CuO$_2$ chemistry is
favored with the possibility of a subdominant incoherent
contribution.
\end{abstract}
\newpage
\section{Introduction}

Carbotte {\it et al.}\cite{carb} recently suggested that a
signature of the $41\,$meV spin resonance which is measured in
spin polarized inelastic  neutron scattering experiments\cite{Bourges}
in the superconducting state of optimally doped YBa$_2$Cu$_3$O$_{6.95}$
(YBCO), is also seen in the in-plane infrared optical
spectrum as a function of energy $\omega$. Following this
initial suggestion Schachinger and Carbotte\cite{schach1,schach2}
proceeded to analyze optical spectra in other high-$T_c$
compounds and found that similar spin resonances develop in
many of the cuprates but not in all. These authors provided
a simple technique\cite{carb,schach1,schach2,mars1,mars2}
to extract from infrared data an approximate electronic
carrier-spin fluctuation spectral density which not only
gives a picture of the position and width of the spin
resonance, but also gives an absolute measure of its coupling to the
charge degrees of freedom. More recently, consideration of in-plane optical
data\cite{schach3} in YBCO at several temperatures
reproduced well the temperature evolution of the $41\,$meV
resonance\cite{dai} measured in neutron scattering experiments
and gave a carrier-spin fluctuation spectral density which showed
considerable variation with temperature. It consists of a broad
spin fluctuation background which extends, in frequency, to
several hundred millivolts and which persists in the normal
state. Superimposed on this background is a spin resonance
contribution which exists only in the superconducting state for
optimally doped YBCO and grows as $T$ is lowered.
 The growth of the resonance
reflects modifications to the spin spectrum connected with the
development of the superconducting state as the temperature
is lowered below $T_c$ (the superconducting critical temperature).
It is found that
feedback effects increase the stability of superconductivity
and, within an Eliashberg formalism for the superconducting state,
lead directly to a ratio of the $d$-wave gap amplitude to the
value of $T_c$ which is considerably larger\cite{gur} than the
BCS value of 4.3 in agreement with experiment.
 
Other observed properties of the superconducting
state can also be understood directly within the same framework. Most
prominent among these are  (1) agreement with experiment
for the value of the zero temperature condensation energy per
copper atom,\cite{loram,normann} (2) the fact that only
about one third of the total optical spectral weight condenses
into the superfluid density at zero temperature.\cite{tanner}
This arises because it is mainly the coherent part
of the charge carrier spectral density which condenses while
the incoherent boson assisted background is largely unaffected
by the condensation. (3) The large peak observed in the microwave
conductivity around $35\,$K\cite{bonn} is understood as due
to the collapse of the electronic scattering rates\cite{nuss}
which results from a low frequency gaping of the spin fluctuation
spectrum due to the onset of superconductivity. (4) There is
also a corresponding peak in the thermal conductivity which the
calculations explain.\cite{matsuk}

In this paper we wish to extend the work to the $c$-axis
optical conductivity\cite{homes,cooper,hirschf} to see if
it too can be satisfactorily understood within a generalized
Eliashberg formalism with kernels determined from in-plane
optical conductivity data.\cite{puchk} The out-of-plane
optical response has been the subject of considerable recent
interest\cite{anders1,anders2,basov,katz,ioffe,kim,chak,dahm,%
winkee} due in part to the observation of an important
violation of the conventional Ferrell-Grover-Tinkham\cite{ferr}
(FGT) sum rule. This sum rule is  obeyed by conventional
superconductors but
not by some high $T_c$ cuprates in the direction perpendicular
to the CuO$_2$ planes. In this case the missing spectral weight
under the real part of the optical conductivity which disappears
on entering the superconducting state, does not necessarily
equal the superfluid density which condenses into a delta
function at zero frequency $\omega = 0$ in the real part of
the optical conductivity.
In this paper we will concentrate on the frequency dependence
of the $c$-axis optical conductivity $\sigma_c(T,\omega)$ at
low temperatures $T$ in the superconducting state for
$\omega > 0$.

\section{Formalism}

The out-of-plane conductivity $\sigma_c(T,\omega)$ at temperature
$T$ and frequency $\omega$ is related to the current-current
correlation function $\Pi_c(T,i\nu_n)$ at the boson Matsubara
frequency $\nu_n = 2n\pi T,\,n = 0,\pm1,\pm2,\ldots$,
analytically continued to real frequency $\omega$,
 and to the $c$-axis kinetic
energy $\langle H_c\rangle$:\cite{kim,radtke}
\begin{equation}
  \label{eq:1}
  \sigma_c(T,\omega) = {1\over\omega}\left[
   \Pi_c(T,i\nu_n\to\omega+i0^+)-e^2d^2\langle H_c\rangle
   \right],
\end{equation}
with $e$ the charge on the electron and $d$ the distance
between planes in the $c$-direction.
In terms of the in-plane thermodynamic Green's function
$\hat{G}({\bf k},i\omega_n)$ and for coherent hopping
$t_\perp({\bf k})$ perpendicular to the CuO$_2$ planes
\begin{mathletters}
  \label{eq:2}
  \begin{equation}
    \label{eq:2a}
   \Pi_c(T,i\nu_n) = 2(ed)^2T\sum\limits_{\omega_m}
   \sum\limits_{\bf k} t^2_\perp({\bf k})\,{\rm tr}\left\{
   \hat{\tau}_0\hat{G}({\bf k},i\omega_m)\hat{\tau}_0
   \hat{G}({\bf k},i\omega_m+i\nu_n)\right\}
  \end{equation}
and
\begin{equation}
  \label{eq:2b}
   \langle H_c\rangle = 2T\sum\limits_{\omega_m}
   \sum\limits_{\bf k} t^2_\perp({\bf k})\,{\rm tr}\left\{
   \hat{\tau}_3\hat{G}({\bf k},i\omega_m)\hat{\tau}_3
   \hat{G}({\bf k},i\omega_m)\right\}  
\end{equation}
\end{mathletters}
In Eqs. (\ref{eq:2}) the $2\times2$ Nambu
Green's function $\hat{G}({\bf k},i\omega_m)$ describes
the in-plane  dynamics of the charge carriers with momentum
{\bf k} in the two dimensional CuO$_2$ plane Brillouin
zone, $i\omega_n = i(2n+1)\pi T$ for temperature $T$,
$n = 0,\pm1,\pm2,\ldots$, and is given by
\begin{equation}
  \label{eq:3}
  \hat{G}({\bf k},i\omega_n) = {i\tilde{\omega}(i\omega_n)\hat{\tau}_0
  + \zeta_{\bf k}\hat{\tau}_3 +\tilde{\Delta}_{\bf k}(i\omega_n)
  \hat{\tau}_1\over -\tilde{\omega}^2(i\omega_n)-\zeta^2_{\bf k}-
  \tilde{\Delta}^2_{\bf k}(i\omega_n)},
\end{equation}
where the $\hat{\tau}$'s are the Pauli $2\times2$ matrices,
$\zeta_{\bf k}$ is the band energy of the charge carriers
as a function of their momentum {\bf k}, $\tilde{\Delta}_{\bf k}%
(i\omega_n)$
is the renormalized gap and $\tilde{\omega}(i\omega_n)$ the renormalized
Matsubara frequencies. In our model these quantities are
determined as solutions of Eliashberg equations suitably
generalized to describe a $d$-wave superconductor in terms
of a charge carrier-spin fluctuation spectral density
$I^2\chi(\omega)$ which we have determined previously from
consideration of the in-plane optical conductivity.\cite{carb,%
schach1,schach2} The method\cite{carb,schach1,schach2,%
mars1,mars2,schach3} is to use the second derivative
${1\over 2\pi}{d^2\over d\omega^2}[\omega\tau^{-1}(\omega)]$ of the
optical scattering rate $\tau^{-1}(\omega)$ in the superconducting
state as a first estimate in the construction of a model for the
underlying charge carrier - spin fluctuation spectral density
$I^2\chi(\omega)$.

In Eqs.~(\ref{eq:2}) the out-of-plane matrix element $t_\perp({\bf k})$
can depend on the in-plane momentum {\bf k}. Models have been summarized
in the recent preprint of Sandeman and Schofield\cite{sand} who
refer to previous literature.\cite{xiang,anders,xiang1} A possible
choice is $t_\perp({\bf k}) = t_\perp$, a constant. But, consideration
of the chemistry of the CuO$_2$ plane and of the overlap of one plane
with the next, suggests a form $t_\perp({\bf k}) = %
\cos^2(2\phi)$ where $\phi$ is the angle of {\bf k} in the
two dimensional CuO$_2$ Brillouin zone for the plane motion. This
matrix element eliminates the contribution from nodal quasiparticles
entirely from the $c$-axis motion.

For incoherent impurity induced $c$-axis charge transfer
Eqs. (\ref{eq:2}) are to be modified. After an impurity
configuration average we obtain
\begin{mathletters}
\label{eq:4}
\begin{eqnarray}
  \Pi_c(T,i\nu_n) &=& 2(ed)^2T\sum\limits_m
   \sum\limits_{{\bf k},{\bf k}'}\overline{V^2_{{\bf k},{\bf k}'}}\left\{
   \hat{\tau}_0\hat{G}({\bf k},i\omega_m)\hat{\tau}_0
   \hat{G}({\bf k},i\omega_m+i\nu_n)\right\}\label{eq:4a}\\
  \langle H_c\rangle &=& 2T\sum\limits_m
   \sum\limits_{{\bf k},{\bf k}'} \overline{V^2_{{\bf k},{\bf k}'}}\left\{
   \hat{\tau}_3\hat{G}({\bf k},i\omega_m)\hat{\tau}_3
   \hat{G}({\bf k},i\omega_m)\right\}\label{eq:4b},
\end{eqnarray}
\end{mathletters}
with $\overline{V^2_{{\bf k},{\bf k}'}}$ the average of the square
of the impurity potential.
 If the impurity potential was taken to conserve momentum,
which it does not, we would recover Eqs. (\ref{eq:2}). Various
models could be taken for $\overline{V^2_{{\bf k},{\bf k}'}}$.
Here we use a form introduced by Kim\cite{kim} and
Hirschfeld {\it et al.}\cite{hirschf}
\begin{equation}
  \label{eq:4c}
  \overline{V^2_{{\bf k},{\bf k}'}} = \vert V_0\vert^2+
   \vert V_1\vert^2\cos(2\phi)\cos(2\phi')
\end{equation}
with $\phi$ and $\phi'$ the directions of {\bf k} and
${\bf k}'$ respectively.

As written Eqs. (\ref{eq:2}) and (\ref{eq:3}) involve Green's
functions in the imaginary frequency Matsubara representation.
To obtain the conductivity $\sigma_c(T,\omega)$ as a function
of real frequency an analytic continuation in Eq. (\ref{eq:1})
to real frequencies is needed. This could be done by Pad\`e
approximates, but here we analytically continue the
entire equations and work with real frequency Eliashberg
equations. For a $d$-wave superconductor with gap
$\Delta_{\bf k} = \Delta_0\cos(2\phi)$ on a cylindrical
Fermi surface where $\phi$ is the polar angle in the two
dimensional CuO$_2$ Brillouin zone, the basic equations
for $\tilde{\Delta}$ and $\tilde{\omega}$
which include inelastic scattering due to a boson exchange
mechanism between the charge carriers and described by the
spectral density $I^2\chi(\omega)$\cite{schach4,schach5}
take on the form
\begin{mathletters}
\label{eq:5}
\begin{eqnarray}
  \lefteqn{\tilde{\Delta}(\nu+i\delta;\phi) = i\pi Tg
  \sum\limits_{m=0}^\infty\cos(2\phi)\left[\lambda(\nu-i\omega_m)+
  \lambda(\nu+i\omega_m)\right]\left\langle
  {\tilde{\Delta}(i\omega_m;\phi')\cos(2\phi')\over
  \sqrt{\tilde{\omega}^2(i\omega_m)+\tilde{\Delta}^2(i\omega_m;
  \phi')}}\right\rangle'}\nonumber\\
 &&+i\pi\int\limits^\infty_{-\infty}\!dz\,\cos(2\phi)
  I^2\chi(z)\left[n(z)+f(z-\nu)\right]\left\langle
  {\tilde{\Delta}(\nu-z+i\delta;\phi')\cos(2\phi')\over
  \sqrt{\tilde{\omega}^2(\nu-z+i\delta)+\tilde{\Delta}^2(\nu-z+i\delta;
  \phi')}}\right\rangle',\label{eq:5a}
\end{eqnarray}
and in the renormalization channel
\begin{eqnarray}
  \tilde{\omega}(\nu+i\delta) &=& \nu+i\pi T%
  \sum\limits_{m=0}^\infty\left[\lambda(\nu-i\omega_m)-
  \lambda(\nu+i\omega_m)\right]\left\langle
  {\tilde{\omega}(i\omega_m)\over
  \sqrt{\tilde{\omega}^2(i\omega_m)+\tilde{\Delta}^2(i\omega_m;
  \phi')}}\right\rangle'\nonumber\\
  &&+i\pi\int\limits^\infty_{-\infty}\!dz\,
   I^2\chi(z)\left[n(z)+f(z-\nu)\right]\nonumber\\
  &&\times\left\langle
  {\tilde{\omega}(\nu-z+i\delta)\over
  \sqrt{\tilde{\omega}^2(\nu-z+i\delta)+\tilde{\Delta}^2(\nu-z+i\delta;
  \phi')}}\right\rangle' + i\pi\Gamma^+{\Omega(\nu)\over
   c^2+D^2(\nu)+\Omega^2(\nu)}
  \label{eq:5b},
\end{eqnarray}
\end{mathletters}
with $\langle\cdots\rangle$ the angular average over $\phi$, and
\begin{equation}
  \label{eq:7}
  \lambda(\nu) = \int\limits^\infty_{-\infty}\!d\Omega\,
   {I^2\chi(\omega)\over\nu-\Omega+i0^+},
\end{equation}
\begin{equation}
  \label{eq:8}
  D(\nu) =  \left\langle{\tilde{\Delta}(\nu+i\delta;\phi)\over
  \sqrt{\tilde{\omega}^2(\nu+i\delta)-\tilde{\Delta}^2(\nu+i\delta;
  \phi)}}\right\rangle,
\end{equation}
\begin{equation}
  \label{eq:9}
  \Omega(\nu) = \left\langle{\tilde{\omega}(\nu+i\delta)\over
  \sqrt{\tilde{\omega}^2(\nu+i\delta)-\tilde{\Delta}^2(\nu+i\delta;
  \phi)}}\right\rangle.
\end{equation}
Equations (\ref{eq:5}) are a set of nonlinear
coupled equations for the renormalized pairing potential $\tilde{\Delta}%
(\nu+i\delta;\phi)$ and the renormalized frequencies
$\tilde{\omega}(\nu+i\delta)$ with the gap
\begin{equation}
  \label{eq:10}
  \Delta(\nu+i\delta;\phi) = \nu\,{\tilde{\Delta}(\nu+i\delta;\phi)%
  \over\tilde{\omega}(\nu+i\delta)},
\end{equation}
or, if the renormalization function $Z(\nu)$ is introduced in the
usual way as $\tilde{\omega}(\nu+i\delta) = \nu Z(\nu)$ then
\begin{equation}
  \label{eq:11}
  \Delta(\nu+i\delta;\phi) = {\tilde{\Delta}(\nu+i\delta;\phi)%
  \over Z(\nu)}.  
\end{equation}
Here, $\nu$ is a real frequency and $\delta$ is a positive
infinitesimal $0^+$. To arrive at these equations a separable (in
the angular part) model was used for the pairing potential. In
the pairing channel it has the form $g\cos(2\phi)I^2\chi(\omega)
\cos(2\phi')$ with $g$ a constant and $I^2\chi(\omega)$ the
pairing spectral density. This leads to a gap proportional to
$\cos(2\phi)$ by arrangement. No other anisotropies are included
and we note that the renormalization channel (\ref{eq:5b}) is
isotropic with the same spectral density $I^2\chi(\omega)$ as
in Eq.~(\ref{eq:5a}) but with no $g$ value. In general a different
form of the spectral density could come into Eqs.~(\ref{eq:5})
but here, for simplicity, we have the same form
but allowed for the possibility that they do not both have the
same magnitude, i.e.: $g$ needs not be equal to one. These
equations are a minimum set and go beyond a BCS approach and
include the inelastic scattering known to be strong in the
cuprate superconductors. In the normal state at $T$ near $T_c$
the inelastic scattering rate varies linearly in $T$, as it
does in our model, and is of the order of a few times $T_c$.

More complicated models which include the possibility of hot
spots\cite{stoj,ioffe1,hlub,marel} could be introduced in our
work but would not alter the
main points we wish to make. In our formalism hot spots can be
introduced by inserting an angular dependent factor in
Eq.~(\ref{eq:5b}) for the renormalized frequencies. Presently,
this quantity is isotropic but we could multiply the spectral
density $I^2\chi(\omega)$ (which enters this equation) by a
factor which increases the scattering in the antinodal as
compared with the nodal direction. This would complicate the
numerical work but goes beyond what we wish to do here. In
any case, recent analysis of angular resolved photoemission
(ARPES) data on the marginal Fermi liquid (MFL)
phenomenology\cite{abrah1,varma,valla,kamen} indicates that the
inelastic part of the quasiparticle scattering in the
cuprates may be quite isotropic.

For the BCS case we get similar formulas on the real axis:\cite{schur}
\begin{mathletters}
\label{eq:bcs}
\begin{eqnarray}
\tilde{\Delta}_d(\nu,\phi) &=& \Delta_0\cos(2\phi)\label{eq:bcsa}\\
\tilde{\Delta}_s(\nu) &=& i\pi t^+\left\langle
  {\tilde{\Delta}_s(\nu)\over\sqrt{\tilde{\omega}^2(\nu)-
  \tilde{\Delta}^2_s(\nu)-\tilde{\Delta}^2_d(\nu,\phi)}}
  \right\rangle\label{eq:bcsc}\\
\tilde{\omega}(\nu) &=& \nu+i\pi t^+\left\langle{
  \tilde{\omega}(\nu)\over\sqrt{\tilde{\omega}^2(\nu)-
  \tilde{\Delta}^2_s(\nu)-
  \tilde{\Delta}^2_d(\nu;\phi)}}\right\rangle,\label{eq:bcsb}
\end{eqnarray}
\end{mathletters}
with impurities treated in the Born limit.

In Eq.~(\ref{eq:5b}) the impurity scattering rate is proportional
to $\Gamma^+$ and enters only the renormalization channel because
we have assumed a pure $d$-wave model for the gap with zero
average over the Fermi surface. This is expected to be the case
in a tetragonal system. The parameter $c$ in the elastic-scattering
part of Eq.~(\ref{eq:5b}) is zero for resonant or unitary scattering
and infinity in the Born approximation, i.e.: weak scattering
limit. In this case the entire impurity term reduces to the
form $i\pi t^+\Omega(\nu)$ with $c$ absorbed into $t^+$. For
intermediate coupling $c$ is finite. The thermal factors appear
in Eqs.~(\ref{eq:5}) through the Bose and
Fermi distribution $n(z)$  and $f(z)$, respectively.

From the solutions of the generalized Eliashberg equations
(\ref{eq:5}) we can construct the Green's function (\ref{eq:3})
analytically continued to the real frequency axis $\omega$.
In this formulation the expression for the
conductivity $\sigma(T,\omega)$ based on equations
(\ref{eq:2}) and (\ref{eq:3}) is lengthy; nevertheless, it
is given here for completeness. For instance,
the in-plane conductivity, after an integration over {\bf k}
has been performed, is given by:
\widetext
\begin{mathletters}
\label{eq:sig}
\begin{eqnarray}
\lefteqn{\sigma_{ab}(\Omega) = {i\over\Omega}\frac{e^2N(0)v^2_F}
  {2}}\nonumber\\
 && \times\left\langle\int\limits_0^\infty\!{\rm d}\nu\,
  {\rm tanh}\left({\nu\over 2T}\right)\frac{1}
  {E(\nu;\phi)+E(\nu+\Omega;\phi)}\left[
  1-N(\nu;\phi)N(\nu+\Omega;\phi)-
  P(\nu;\phi)P(\nu+\Omega;\phi)\right]\right.
  \nonumber\\
 &&\left.+\int\limits_0^\infty\!{\rm d}\nu\,
  {\rm tanh}\left({\nu+\Omega\over 2T}\right)\frac{1}
  {E^\star(\nu;\phi)+E^\star(\nu+\Omega;\phi)}
  \left[1-N^\star(\nu;\phi) N^\star(\nu+\Omega;\phi)
  \right.\right.\nonumber\\
 &&\left.\left.
  -P^\star(\nu;\phi) P^\star(\nu+\Omega;\phi)
  \right]
  +\int\limits_0^\infty\!{\rm d}\nu\,\left[{\rm tanh}
  \left({\nu+\Omega\over 2T}\right)-{\rm tanh}\left({
  \nu\over 2T}\right)\right]\right.\nonumber\\
 &&\left.\times\frac{1}
  {E(\nu+\Omega;\phi)-E^\star(\nu;\phi)}
  \left[1+N^\star(\nu;\phi) N(\nu+\Omega;\phi)
  +P^\star(\nu;\phi) P(\nu+\Omega;\phi)
  \right]\right.\nonumber\\
 &&\left.+\int\limits_{-\Omega}^0\!{\rm d}\nu\,
  {\rm tanh}\left({\nu+\Omega\over 2T}\right)\left\{
  \frac{1}
  {E^\star(\nu;\phi)+E^\star(\nu+\Omega;\phi)}
  \left[1-N^\star(\nu;\phi) N^\star(\nu+\Omega;\phi)
  \right.\right.\right.\nonumber\\
 &&\left.\left.\left.
   -P^\star(\nu;\phi) P^\star(\nu+\Omega;\phi)
  \right]\right.\right.\nonumber\\
 &&\left.\left.
   +\frac{1}
  {E(\nu+\Omega;\phi)-E^\star(\nu;\phi)}
  \left[1+N^\star(\nu;\phi) N(\nu+\Omega;\phi)
  +P^\star(\nu;\phi) P(\nu+\Omega;\phi)
  \right]\right\}\right\rangle  \label{eq:siga}
\end{eqnarray}
with
\begin{equation}
  E(\omega;\phi) = \sqrt{\tilde{\omega}^2_{\bf k}
   (\omega)-\tilde{\Delta}^2_{\bf k}
   (\omega)} \label{eq:sigb}
\end{equation}
and
\begin{equation}
  N(\omega;\phi) = {\tilde{\omega}_{\bf k}(\omega)\over
   E(\omega;\phi)},\qquad
  P(\omega;\phi) = {\tilde{\Delta}_{\bf k}(\omega)\over
   E(\omega;\phi)}. \label{eq:sigc}
\end{equation}
\end{mathletters}
In the above, $\langle\cdots\rangle$ means an average over the
angle $\phi$ and the star refers to the complex conjugate.
For the $c$-axis conductivity $\langle\cdots\rangle$ is to
be replaced by $\langle\cdots\cos^4(2\phi)\rangle$ and the
prefactor is different with $v_F^2/2$ to be replaced by
$d^2 t_\perp^2$. The above set of equations is valid for
the real and imaginary part of the conductivity as a function
of frequency $\Omega$. It contains only the paramagnetic
contribution to the conductivity but this is fine since we will
be interested in this paper mainly in the real part of the
conductivity for which the diamagnetic contribution is zero.

The real part of the incoherent conductivity along the $c$-axis is in
turn given by (normalized to its normal state value
$\sigma_{1cn}$):\cite{hirschf}
\begin{equation}
{\sigma_{1c}(\Omega)\over\sigma_{1cn}} =
  {1\over\nu}\int\!d\omega\,\left[f(\omega)-f(\omega+\Omega)
  \right]\left[N(\omega+\Omega)N(\omega)
 +\left\vert{V_1\over V_0}\right\vert
   P(\omega+\Omega)P(\omega)\right].\label{eq:inc}
\end{equation}

\section{Results and Discussion}

We begin our discussion by highlighting some of the important
results for the real part of the in-plane conductivity
$\sigma_1(T,\omega)$ as a function of frequency at low
temperatures $T$. We will want to compare our $c$-axis
results with these in-plane results.
 In Fig.~\ref{f1} we show $\sigma_1(\omega)$
vs. $\omega$ within BCS theory for a conventional $s$-wave
gap (solid curve) and compare with the less known results
for a $d$-wave superconducting gap function (dashed curve).\cite{schur}
The solid curve is zero for all frequencies $\omega\le2\Delta_0$
(twice the $s$-wave gap value of $\Delta_0$). This is a well
known result which has its origin in the fact that, at zero
temperature, all the carriers have condensed into the
superfluid and for there to be absorption, two quasiparticle
excitations corresponding to the breaking of Cooper pairs
out of the condensate are needed. This costs $2\Delta_0$ of
energy. In the normal state (dotted curve) the equivalent excitations
correspond to the creation of a hole and a particle (2 excitations).
At $\omega = 2\Delta_0$ there is a sharp absorption edge and the
conductivity jumps to a value close to its normal state
Drude value (dotted line in Fig.~\ref{f1}). In the above,
the gap has been set at $24\,$meV, a value characteristic of
optimally doped YBCO. Also for the normal state Drude
(dotted curve) the impurity scattering rate was
$t^+ = 0.001\,$meV. This very small value was employed so as
to be able to resolve fine structures in the $d$-wave results
which we will now describe.

The dashed line is for a $d$-wave superconductor in the BCS
limit. In the two dimensional CuO$_2$
Brillouin zone the angle $\phi$ gives the direction of
momentum {\bf k} and $\Delta_{\bf k} = \Delta_0\cos(2\phi)$
in our simple model. It is clear that, entering into the
superconducting state reduces $\sigma_1(\omega)$ in the
region of the gap $\Delta_0$ but does not make it go to
zero because even at $T=0$ some of the electrons have zero
or very small gap values, namely those on the main diagonals
of the Brillouin zone for $\phi = \pm\pi/4$. It is also
worth emphasizing that, above $2\Delta_0$ the conductivity goes
back to its normal state value faster than it does for the
$s$-wave case. For $s$-wave the curve for $\sigma_1(\omega)$
is above the normal state Drude for $\omega \ge 2\Delta_0$
and merges with the Drude form from above. For $d$-wave the superconducting
$\sigma_1(\omega)$ is slightly below the Drude in the same
region but not by very much.
 In both cases the missing optical spectral
weight on entering the superconducting state, of course,
goes into the superfluid which provides a delta function response
in $\sigma_1(\omega)$ at $\omega=0$. This is not shown
in our figure. Note also that the $d$-wave conductivity
has a small structure (on the scale of the figure) at
$\omega = \Delta_0$ corresponding to a maximum in slope
for $\sigma_1(\omega)$ and has another small structure at
$\omega = 2\Delta_0$ corresponding to a kink in the curve.
These structures have their origin in the particularities
of the $d$-wave state. For example, in $d$-wave,
the quasiparticle density of states as a function of energy
is linear in $\omega$ at small $\omega$ and has a logarithmic
singularity at $\omega = \Delta_0$ before taking on its
normal state value at large $\omega$. This singularity is
much weaker than for the corresponding $s$-wave case for
 which the density
of states is zero up to $\omega = \Delta_0$ at which point it
displays a inverse square root singularity. This square root
singularity does not
imply a singularity in $\sigma_1(\omega)$ however. As we
have seen in Fig.~\ref{f1}, there is instead only a sharp edge in 
$\sigma_1(\omega)$ at twice the gap value. We can therefore
 expect much smaller
structures in $\sigma_1(\omega)$ at $\omega = \Delta_0$ and
$2\Delta_0$ in $d$-wave than in the $s$-wave case at $2\Delta_0$ and
this is confirmed by
our numerical results given in Fig.~\ref{f1}.

These results
are presented here mainly for comparison with full Eliashberg
results based on the formalism given in the pervious
section which includes inelastic effects.  These can be
very large in the cuprates, and are described by the charge carrier-%
mediating-boson spectral density $I^2\chi(\omega)$ introduced
in the previous section. Several different theoretical results
are presented in Fig.~\ref{f2}. Four different curves are
shown for comparison with each other. They were chosen to
illustrate the main features present in the in-plane
conductivity when inelastic scattering is included. The
gray solid line with a maximum before $\omega = 50\,$meV are BCS
results as in the previous figure. It is totally different
from the other curves and disagree strongly with
experiment (solid curve). The other curves are all theoretical
and based on the Eliashberg equations but with different
models for the charge carrier-mediating-boson spectral
density $I^2\chi(\omega)$. For the dash-dotted curve referred
to as MMP with $\omega_{\rm SF} = 20\,$meV, we employed
the simple spectrum for the interaction with the spin
fluctuations first introduced by Pines and coworkers.%
\cite{millis,mont} It has the form:\cite{schach4,millis,%
mont}
\begin{equation}
  \label{eq:12}
  I^2\chi(\omega) = I^2{\omega/\omega_{\rm SF}\over
      1 + (\omega/\omega_{\rm SF})^2},
\end{equation}
where $I^2$ is a coupling of the carriers to the spin
fluctuations and $\omega_{\rm SF}$ is a characteristic
spin fluctuation frequency. In our work $I^2$ and
$\omega_{SF}$ are adjusted to get the best possible fit
to the normal state conductivity. The same spectrum is
used at all temperatures in the superconducting state
with the value of $g$ in the $\tilde{\Delta}$-channel
(\ref{eq:5a}) of the linearized Eliashberg equations
set to get the measured value of $T_c = 92.4\,$K. 
There are
no other parameters with $g = 0.98$ in Eqs.~(\ref{eq:5}).
 The resulting dash-dotted curve
is to be compared with the BCS result. It is seen to be
quite different. In particular there is no trace of
the small structures found in the BCS $\sigma_1(\omega)$
at $\omega = \Delta_0$ and $2\Delta_0$. In the MMP model
the inelastic scattering at such energies is large and
smears out any such small structures. However,
now there is a significant structure at $\Delta_0$ plus the
energy of the peak in $I^2\chi(\omega)$ namely
$\omega_{\rm SF} = 20\,$meV. Since in a $d$-wave superconductor
electrons at the Fermi energy with momentum along the main
diagonals have zero gap, the boson assisted processes start,
in principle, at zero energy plus the boson energy $(\omega_B)$
rather than twice the gap plus $\omega_B$. But such processes
make only a small contribution to the real part of the
conductivity and are not seen as a prominent structure in this
quantity. At the gap energy $\Delta_0$ the
density of electronic states, however, has a logarithmic
van Hove singularity and this is sufficient to produce the
structure at $\Delta_0+\omega_B$ described above. 
This is discussed in more detail in the
work of Carbotte and coworkers\cite{carb,schach1,schach2,%
mars1,mars2,schach3} and we return to it below. It is
sufficient here to state that a reasonable picture of
the underlying $I^2\chi(\omega)$ can be obtained from a
second derivative of the corresponding optical scattering
rates as a function of energy. This quantity contains a recognizable
picture of $I^2\chi(\omega)$. Here we point out that
after the main Drude peak has largely decayed with increasing
$\omega$, there is a region of low conductivity in
$\sigma_1(\omega)$ which is followed by a rise at higher
energies. This rise corresponds to the boson assisted
incoherent part of the electron quasiparticle spectral
density. The coherent delta function which is also
present in the quasiparticle spectral density
 contributes the Drude. In our work this
part contains about 25\% of the total spectral weight.

The boson assisted region at higher energies can be used
to model $I^2\chi(\omega)$ rather than taking it from some
simple theory such as MMP. We can use the solid curve
which shows the in-plane data\cite{schach3} to get an experimentally
measured $I^2\chi(\omega)$. When this is done 
we find that it consists
of the $41\,$meV spin resonance measured in inelastic
spin polarized neutron scattering experiments, plus a
background extending up to $400\,$meV which we model by
an MMP spectrum (\ref{eq:12}). The resulting $I^2\chi(\omega)$
gives the dashed curve which agrees with the
experimental data in the boson region. This good agreement provides 
strong evidence for charge coupling to
the spin resonance at $41\,$meV which exists only in the
superconducting state of optimally doped YBCO\cite{Bourges} below
$T_c$. The calculations were performed at
$T = 10\,$K which is close enough to zero. The spectral
weight of the spin resonance is largest at $T=0$ and
decreases with increasing $T$ to vanish at $T=T_c$. 
The temperature dependence of the resonance, measured by
neutrons\cite{dai} is also well represented in the
optical data.\cite{schach3} There are two other
sets of results presented in Fig.~\ref{f2}. The dotted curve
uses $I^2\chi(\omega)$ in the calculations but also includes
impurities in Born scattering with $t^+ = 0.32\,$meV. This
does not much improve the agreement between theory and
experiment in the frequency region below the boson assisted region.
The dash-double-dotted curve also includes impurities but
this time the unitary limit is used with $\Gamma^+ = 0.63\,$meV.
We see that we now have an excellent fit to the data
throughout the entire frequency range. We note the impurities
hardly affect the boson assisted region which determines
$I^2\chi(\omega)$ but dominate at low $\omega$ and that
some resonant scattering must be included to get agreement
in this region.

The agreement obtained between Eliashberg theory and
experiment is not limited to the real part of $\sigma(\omega)$.
In Fig.~\ref{f2a} we show results for the imaginary part
of $\sigma(\omega)$, more precisely for $\omega\sigma_2(\omega)$.
The curves are labeled as in Fig.~\ref{f2}. The solid line
is the data which is not well represented by the dash-dotted
curve based on an MMP model for the spectral density. To
get agreement it is necessary to include in an Eliashberg calculation
the $41\,$meV resonance as we have done for the real part of
$\sigma(\omega)$. The pure case without impurity scattering
fits the boson assisted region as well as the other two
curves with Born scattering $t^+ = 0.32\,$meV (dotted curve)
and resonant scattering $\Gamma^+ = 0.63\,$meV (dash-double-dotted
curve). However, at lower frequencies, below $25\,$meV, only
the curve with resonant scattering fits the data well.
Returning to the phonon assisted region the large dip in
$\omega\sigma_2(\omega)$ seen around $75$-$80\,$meV and
reproduced by our theory is a signature of the $41\,$meV
resonance in this quantity. The dip is not present in a theory
of the infrared optical conductivity based on $d$-wave BCS
theory.\cite{schur} In such a theory all structures in
$\omega\sigma_2(\omega)$ vs. $\omega$ fall around or below
twice the value of the gap amplitude $\Delta_0$. The
structure observed in the data at approximately $2\Delta_0$
plus the resonance energy $\omega_r$ and is thus the signature
of $\omega_r$ in $\omega\sigma_2(\omega)$. For the MMP
model (dash-dotted curve) there is a dip too but it falls at
the wrong energy.

We give a few more details about the second derivative technique
which was used by us to construct the underlying spectral density
$I^2\chi(\omega)$ from infrared data. A function $W(\omega)$
\begin{equation}
  \label{eq:wo}
  W(\omega) = {1\over 2\pi}{d^2\over d\omega^2}\left [\omega\over
              \tau(\omega)\right]
\end{equation}
is constructed as a guide only. In the normal state and at
low temperatures this function is almost exactly\cite{carb,%
schach1,schach2,mars1,mars2,schach3} equal to the input
$I^2\chi(\omega)$ for models based on the nearly antiferromagnetic
Fermi liquid (NAFFL). Of course, $I^2\chi(\omega)$ is seen in
$W(\omega)$ through electronic processes. But in the normal
state the electronic density of states $N(\varepsilon)$ is
constant and so does not lead to additional structures in
$W(\omega)$ that are not in $I^2\chi(\omega)$ which would then corrupt
the signal, if the aim is to obtain $I^2\chi(\omega)$ from $W(\omega)$.
This is no longer the case in the superconducting
state because of the logarithmic van Hove singularities in
$N(\varepsilon)$ and these do indeed strongly influence the
shape of $W(\omega)$ and introduce additional structures in
$W(\omega)$ corresponding
to combinations of the positions of the singularity in
$N(\varepsilon)$ and the peak in $I^2\chi(\omega)$ at
$\omega_r$ (resonance frequency) as described
by Abanov {\it et al.}\cite{abanov} The structures in $W(\omega)$
corresponding to these singularities contaminate the signal
in the sense that $W(\omega)$ in the superconducting state is no longer
equal to the input $I^2\chi(\omega)$.\cite{schach1,schach2}
In fact, only the resonance peak appears clearly at $\Delta_0+%
\omega_r$ and its size in $W(\omega)$ is about twice the value
of $I^2\chi(\omega)$ at that frequency. In some cases the
tails in $W(\omega)$ also match well the tails in $I^2\chi(\omega)$.
In the end, of course, $W(\omega)$ serves only as a guide and it
is the quality of the final fit to the conductivity data that
determines the quality of the derived $I^2\chi(\omega)$.

Nevertheless, besides giving a measure of the coupling of the
charge carriers to the spin resonance $W(\omega)$ can also be
used to see the position of density of states singularities, as shown in
Fig.~\ref{f2b} where $I^2\chi(\omega)$ (gray squares) and
$W(\omega)$ (solid line) derived from our theoretical results
are compared. Also shown by vertical arrows are the positions
of $\Delta_0+\omega_r$, $2\Delta_0+\omega_r$,
$\Delta_0+2\omega_r$, and $2\Delta_0+2\omega_r$. We note
structures at each of these places  and this information is
valuable. Note that at $2\Delta_0+\omega_r$ the large negative
oscillation seen in $W(\omega)$ is mainly caused by the kink
in $I^2\chi(\omega)$ (gray squares) at about $55\,$meV.
 The density of electronic states effects clearly
distorted the spectrum above the resonance peak
and $W(\omega)$ stops agreeing with the input $I^2\chi(\omega)$
in this region until about $150\,$meV where agreement is
recovered. In summary, $W(\omega)$ contains some information
on singularities in $N(\varepsilon)$ as well as on the shape and size of
$I^2\chi(\omega)$ and, in the superconducting state, the two effects
cannot be clearly separated. Nevertheless, $W(\omega)$
remains a valuable intermediate step in the construction of
a charge carrier-exchange boson
interaction spectral density from optical data. 

We next turn to the $c$-axis dynamics and present results
first for a BCS model. To proceed we need to specify the
transverse coupling. We have presented in the theory
section two possible models. The first one is coherent
tunneling with the matrix element $t_\perp({\bf k})$
conserving in-plane momentum and probably equal to
$t_\perp\cos^2(2\phi)$ where $\phi$ is the angle of {\bf k}
in the two dimensional CuO$_2$ Brillouin zone. The
other model is incoherent tunneling for which momentum is
not conserved. To be definite we will use
$\vert V_1/V_0\vert = 1$ (see Eq.~(\ref{eq:4c})) in the
impurity potential. Other values of $\vert V_1/V_0\vert$
have been studied by Hirschfeld {\it et al.}\cite{hirschf}
and by others\cite{kim,winkee} to which the reader is referred for
more details.

In Fig.~\ref{f3} we show BCS results for the real part of the
$c$-axis conductivity $\sigma_{1c}(\omega)$
vs. $\omega$ and compare with the in-plane conductivity
$\sigma_1(\omega)$ (dotted curve). In this case the gap is
$24\,$meV and the in-plane impurity scattering rate is
$t^+ = 0.1\,$meV. We see that the structure at $\Delta_0$ and
$2\Delta_0$
in the dotted curve for $\sigma_1(\omega)$ have been
smeared out somewhat more when compared with the equivalent results
shown in Fig.~\ref{f1}. This is due to the larger impurity
content. This curve is for comparison with our $c$-axis
results which we now describe. The solid curve is for the
coherent tunneling case
with $t_\perp({\bf k}) = t_\perp\cos^2(2\phi)$. This hopping probability
eliminates the nodal quasiparticles along $(\pi,\pi)$ which
do not participate in the out-of-plane dynamics.
This gets rid of much of the remaining Drude like
contribution at very low $\omega$ which is still clearly
present in the dotted curve although it is substantially
reduced by superconductivity as compared to the normal state
Drude (see Fig.~\ref{f1}). The solid curve is small at small
$\omega$ and peaks just below $\omega \simeq 2\Delta_0$
where the dotted curve for the in-plane case has a small
structure. At higher energies, there is little difference
between the in-plane and out-of-plane $\sigma_1(\omega)$
although the magnitude of these two quantities is of course
very different. If instead of using $t_\perp({\bf k})$ we had used
a constant $t_\perp$ for the $c$-axis transport,
the out-of-plane $\sigma_{1c}(\omega)$
would mirror the in-plane case $\sigma_1(\omega)$.
This would also hold in the more complex Eliashberg
calculations. Only the magnitude is different between in-plane
and out-of-plane in this case because of differences in
over all multiplicative factors in front of the expression
for the conductivity. The final
curve in Fig.~\ref{f3}, dash-dotted, is for the incoherent
case. We see that it too is near zero at small $\omega$
although it rises out of zero more rapidly than does the solid
curve. It shows no structure whatever at the gap or at twice
its value. The main rise is accomplished within the region
$\omega \stackrel{<}{\sim} 2\Delta_0$. At high $\omega$
it saturates to a constant value of one. This is because
we have normalized our results to the normal state conductivity
and the curve becomes very flat. The corresponding normal
state conductivity would be a horizontal line at this
saturated value, constant for all $\omega$. This behavior
bares no relation to the in-plane coherent result.

The results for $\sigma_{1c}(\omega)$ presented so far are
for comparison with those based on solutions for the
Eliashberg equations given in the previous section and which
properly include inelastic scattering through the spectral
density $I^2\chi(\omega)$. Before presenting our $c$-axis results
in this case we stress again that the boson exchange
kernel $I^2\chi(\omega)$ is an
in-plane quantity and is taken from our discussion of the in-plane
conductivity. It is not fitted to any $c$-axis data. It is
to be used unchanged to calculate the out-of-plane
conductivity assuming coherent hopping with
$t_\perp({\bf k}) = t_\perp\cos^2(\phi)$. The solid curve
in Fig.~\ref{f4} are the
in-plane Eliashberg results which are included for comparison with the
dashed curve which is for the $c$-axis. In the boson assisted
region, which would not exist in a BCS theory, both curves
have a remarkably similar behavior. At very low frequencies,
a region which comes mainly from the coherent delta function part of
the carrier spectral density, and which is the only part included in
BCS, we note a narrow Drude-like peak in the solid curve.
This part is suppressed in the $c$-direction
(dashed curve) because the contribution from the nodal
quasiparticles are effectively left out by the
$t_\perp\cos^2(\phi)$ weighting term. Also, shown for comparison
are our previous BCS results for coherent hopping (dotted
curve). These results show no resemblance to our
Eliashberg results and also do not agree with
experiment. What determines the main rise in the region
beyond the Drude part of the conductivity in $\sigma_{1c}(\omega)$
are the boson assisted processes and this rise does not signal
the value of the gap or twice the gap for that matter but rather
a combination of $\Delta_0$ and the resonance energy $\omega_r$.

In Fig.~\ref{f5} we compare the data from Homes {\it et
al.}\cite{homes} on the same graph for in-plane (dotted)
and out-of-plane (solid) conductivity $\sigma_1(\omega)$.
It is clear that in the $c$-direction, the nodal
quasiparticle seen in the dotted curve are strongly
suppressed. This favors the $t_\perp\cos^2(2\phi)$
matrix element for the $c$-axis dynamics as we have just seen.
Further, in the
boson assisted region the two curves show almost perfect
agreement with each other, which again favors the $t_\perp\cos^2(2\phi)$
coupling as illustrated in the theoretical curves of
Fig.~\ref{f4}. One difference is that the main rise,
indicating the onset of the boson assisted incoherent
(in-plane) processes, appears to have shifted slightly
toward lower
frequencies in the $c$-axis data as opposed to a shift
to slightly higher frequencies in our theory. It should
be remembered, however, that in the raw $c$-axis data,
large structures appear in the conductivity due to
direct phonon absorption and these need to be subtracted
out, before data for the electronic background of Fig.~\ref{f5}
can be obtained. In view of this, it is not clear to us how
seriously we
should take the relatively small disagreements that we
have just described between theory and experiment.

\begin{figure}
\caption{Comparison with the data of Homes {\it et al.}%
\cite{homes} for the $c$-axis conductivity (black solid curve).
The theoretical curves were obtained in a BCS theory,
solid gray (coherent), dotted gray (incoherent) and the
others in Eliashberg theory with MMP model and impurities
$t^+ = 0.311\,$meV. The black dotted curve is for incoherent $c$-axis
with $\vert V_1/V_0\vert = 1$, the dashed for coherent
$c$-axis with $t_\perp(\phi) = t_\perp\cos^2(2\phi)$
with $\phi$ an angle in the two dimensional CuO$_2$
Brillouin zone, and the dash dotted is a fit to the data
provided by a mixture of coherent and incoherent. We
stress that this last fit is for illustrative purposes only,
and is not unique.}
\label{f6}
\end{figure}
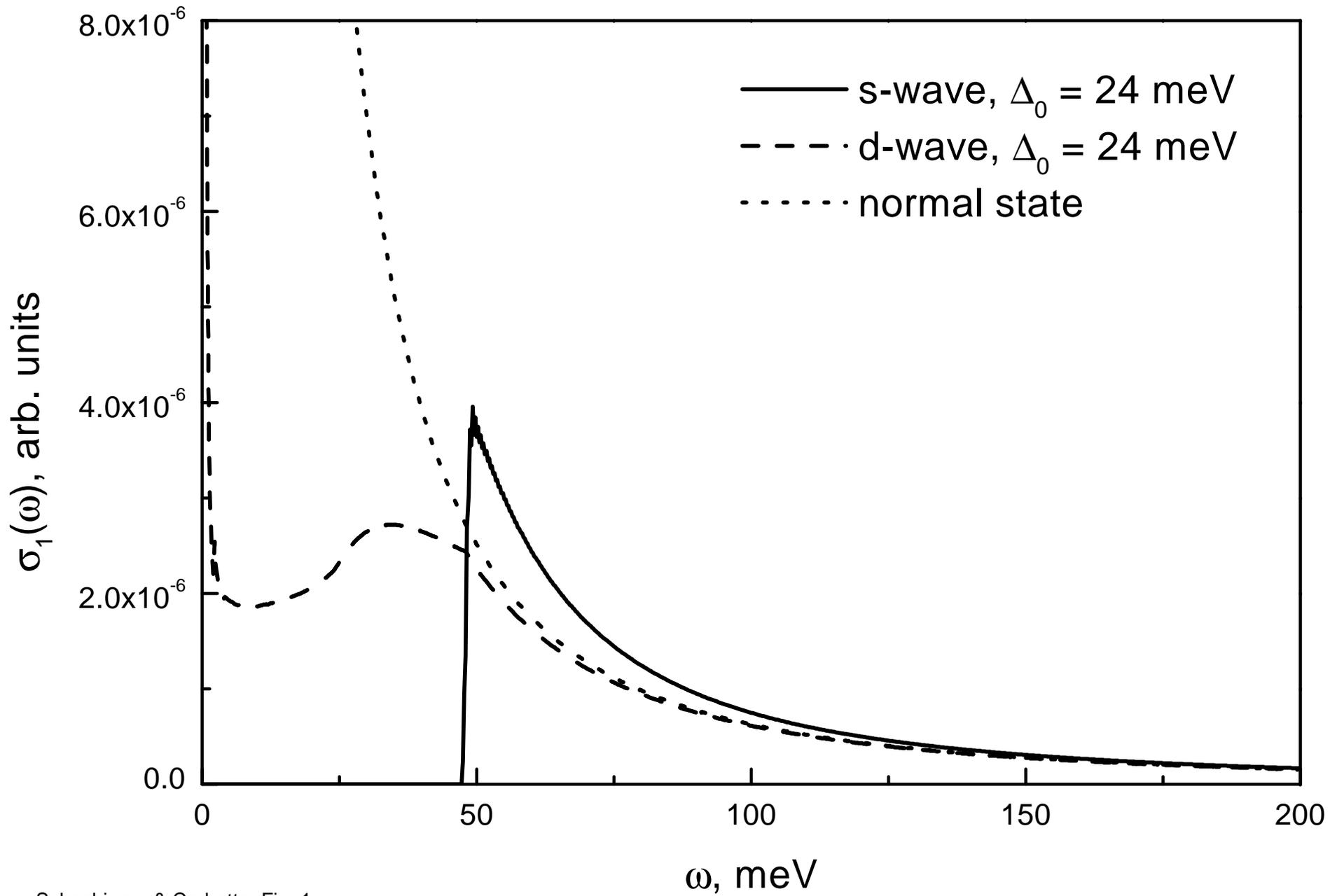
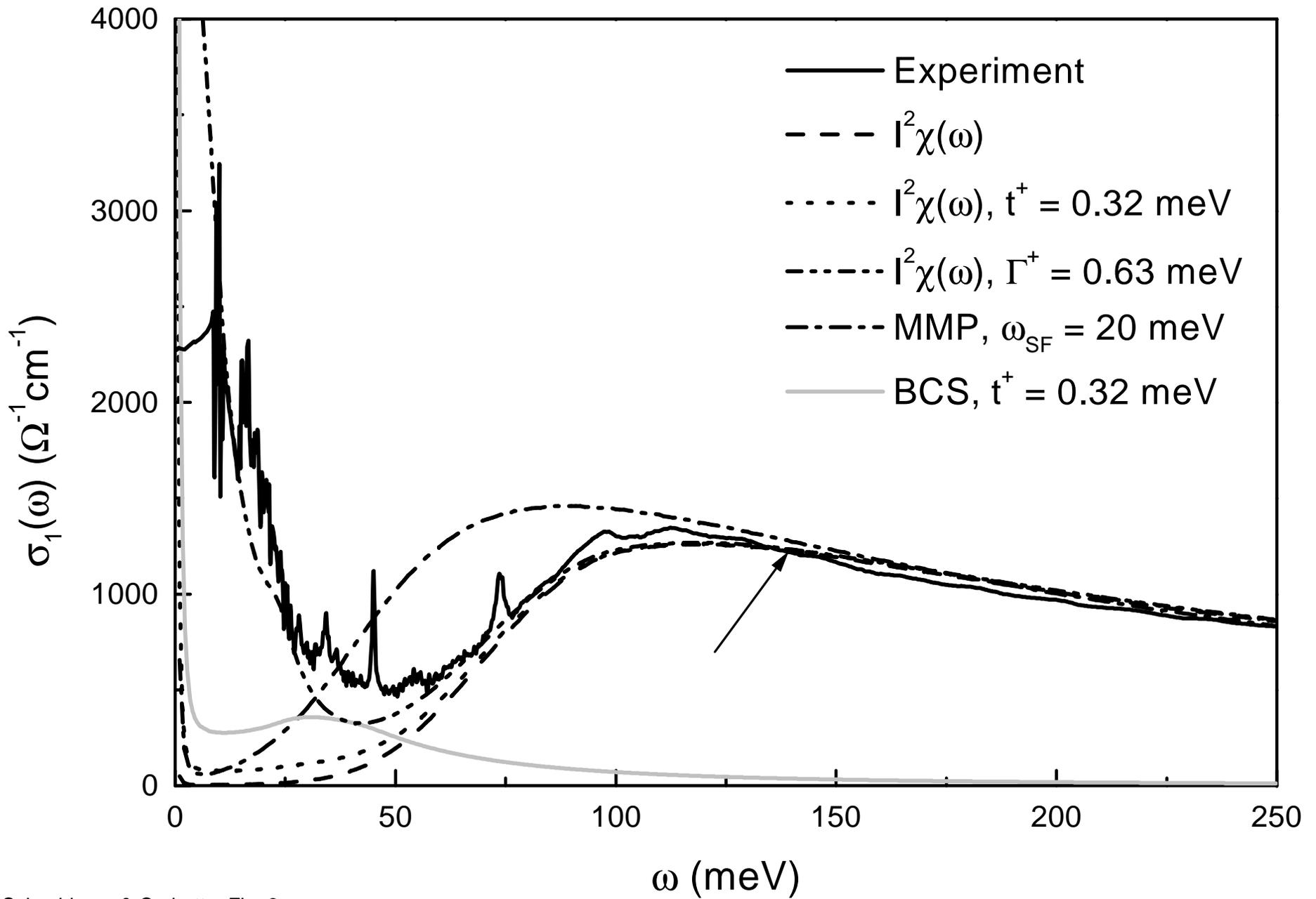
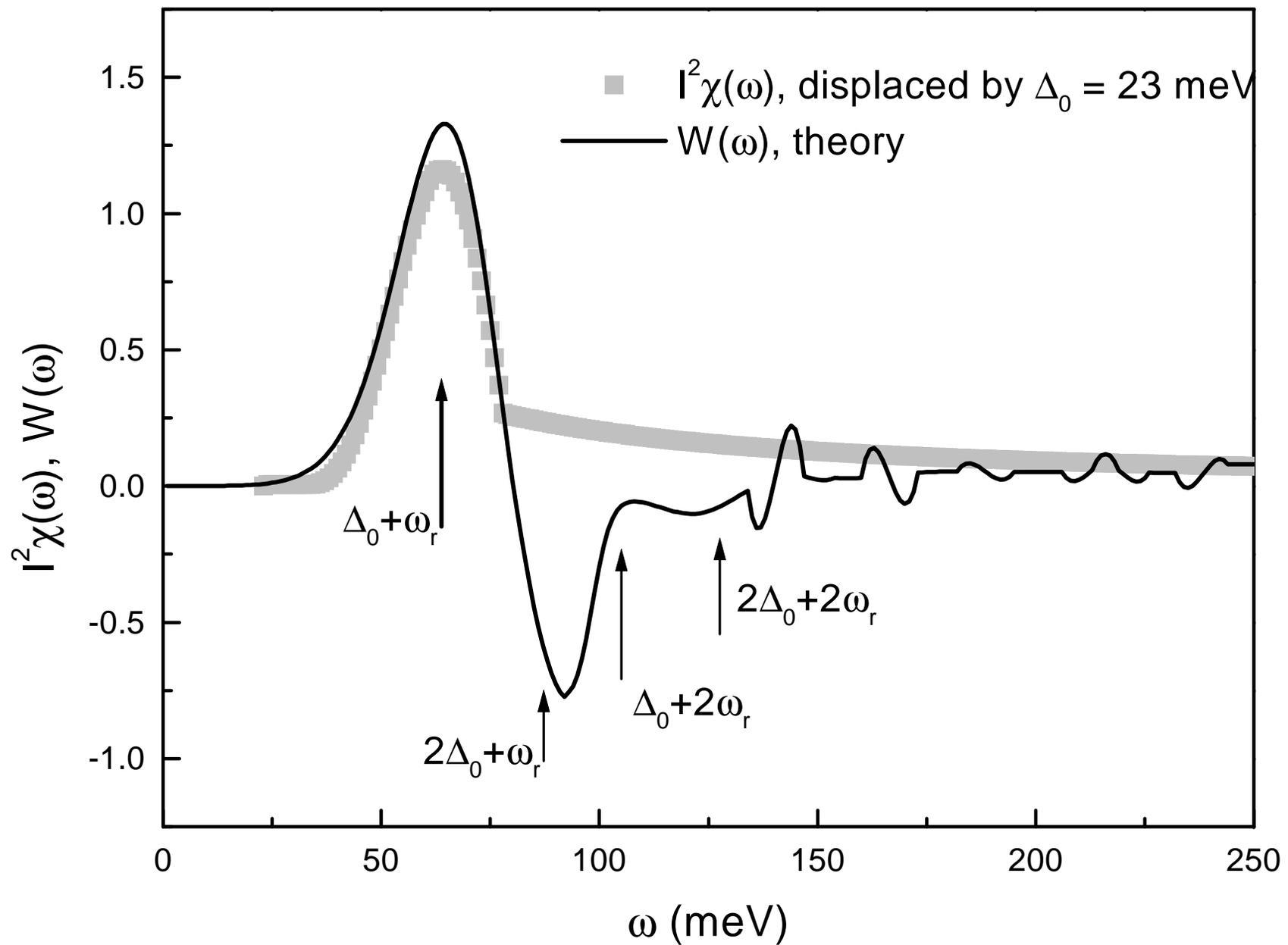
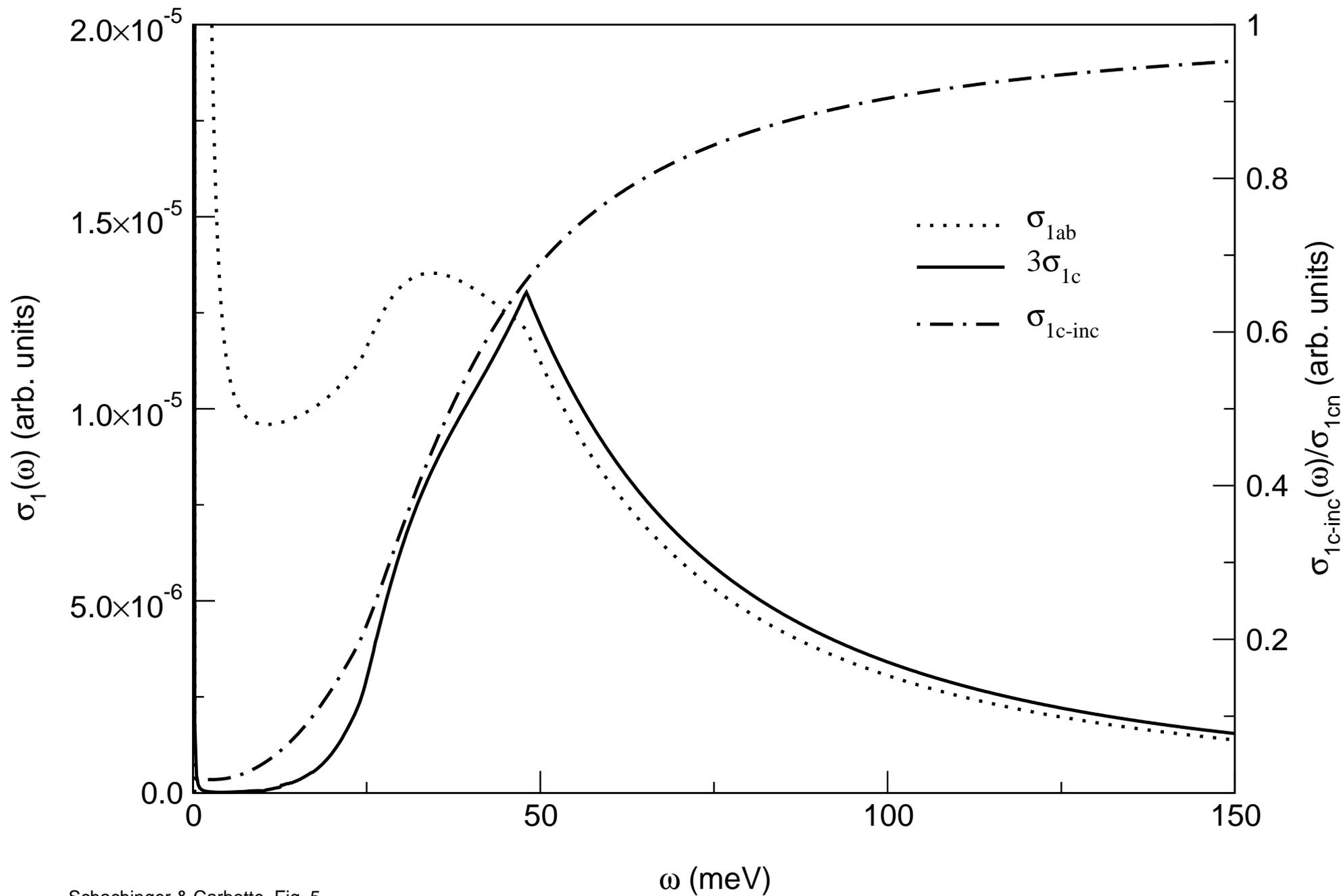
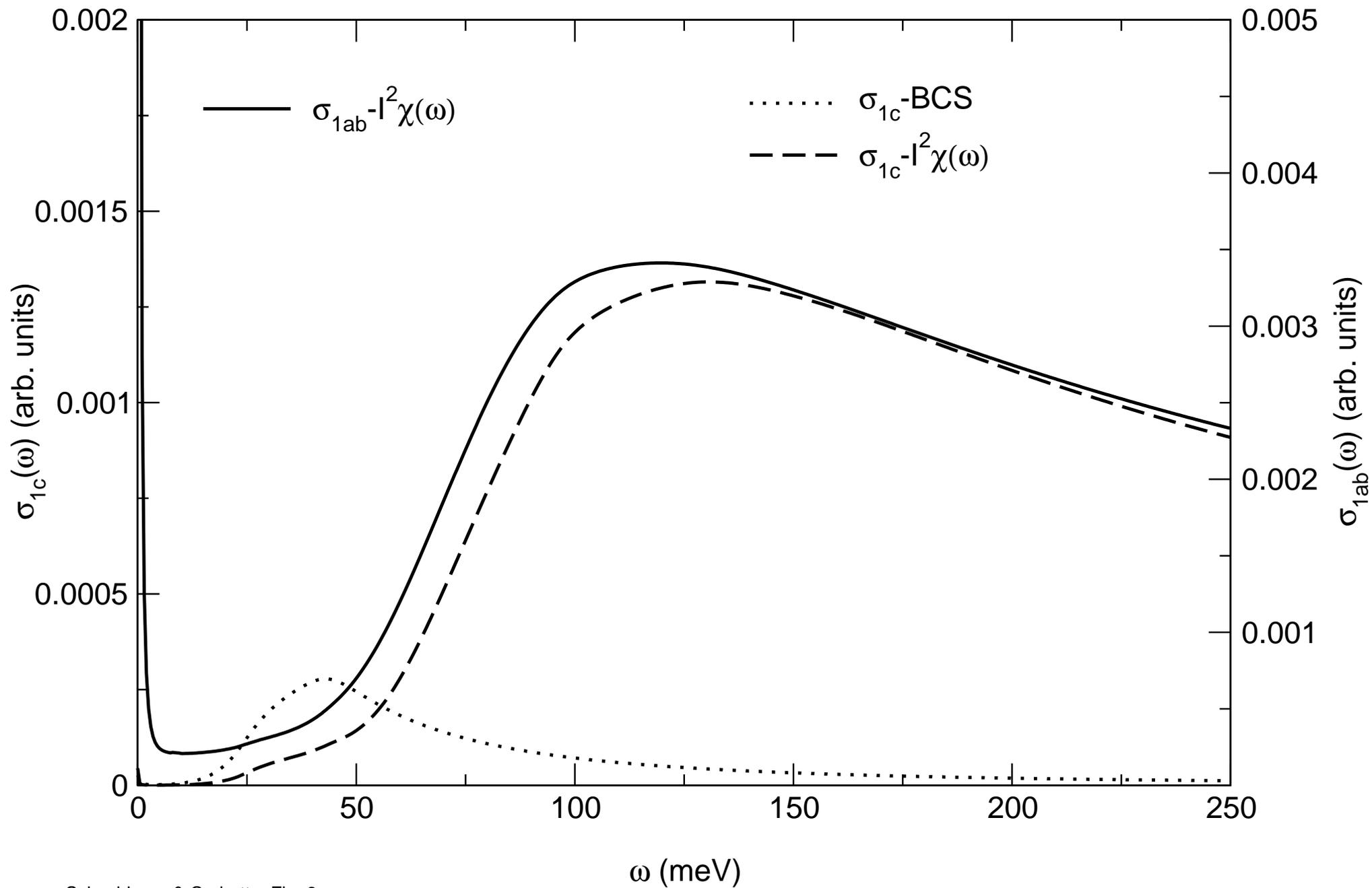
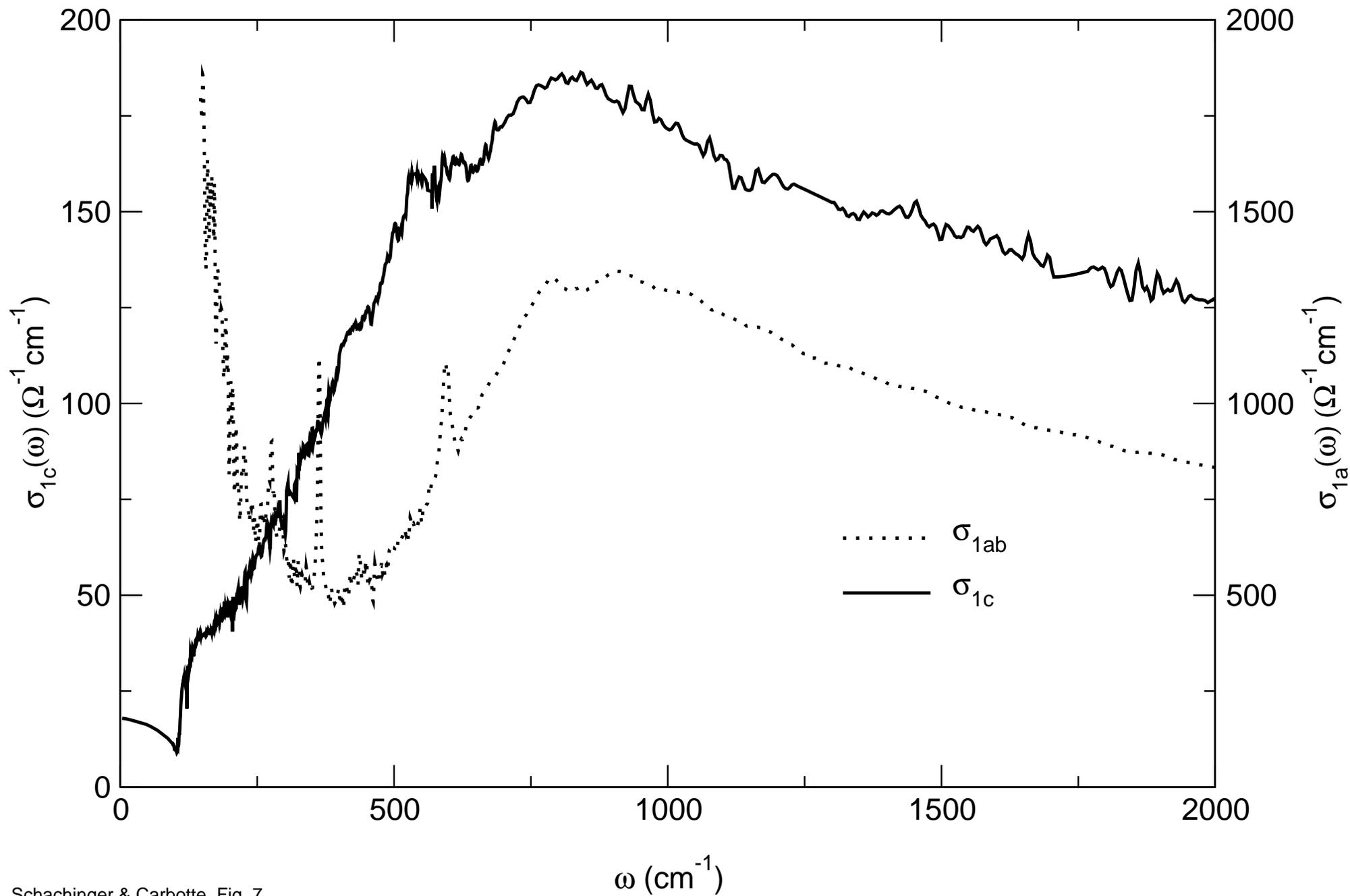

With the above
reservation kept in mind we show in our last Fig.~%
\ref{f6} a comparison of various theoretical results with
experimental $c$-axis conductivity (black solid
line). There are five additional curves. The black ones
are obtained from an Eliashberg calculation based on the
MMP model for $I^2\chi(\omega)$ with impurities $t^+ =%
0.32\,$meV included to simulate the fact that the samples
used are not perfect i.e. are not completely pure, but this
parameter does not play a critical role in our discussion.
Incoherent $c$-axis coupling is assumed with
$\vert V_1/V_0\vert = 1$ (black dotted). It is clear that this
curve does not agree with the data and that the coupling
along $c$ cannot be dominated by incoherent hopping
between planes. This is also in agreement with the results
of a theoretical study by Dahm {\it et al.}\cite{dahm}
who also observed better agreement for coherent $c$-axis
conductivity in the overdoped regime.
On the other hand the fit to the black dashed line is good in comparison.
It uses the same MMP model but with
coherent coupling of the form $t_\perp({\bf k}) = t_\perp\cos^2(2\phi)$.
This fit may already be judged satisfactorily but it
should be remembered that if we had used the model of
$I^2\chi(\omega)$ with the $41\,$meV peak included instead
of MMP, the agreement would have deteriorated.
This is troubling since one
would expect that coupling to the $41\,$meV spin
resonance would be stronger in the $c$-direction data than
it is in the in-plane data. This is because the $c$-axis
emphasizes the hot spots around the antinodal directions
which connect best to $(\pi,\pi)$ in the magnetic susceptibility.
This is the position in momentum space where this spin
resonance is seen to be located in optimally doped YBCO.
On the other hand, recent ARPES data\cite{abrah1,varma,valla,%
kamen} which fit well the MFL (marginal Fermi liquid)
phenomenology show little
in-plane anisotropy for scattering around the Fermi surface
and this is consistent with the findings here.

The dash-dotted curve in Fig.~%
\ref{f6} illustrates the fit to the data that can be achieved with a
dominant coherent piece and subdominant incoherent
contribution. It is not clear to us that such a close fit
is significant given the uncertainties in the data and
the lack of uniqueness in the fitting procedure. It does, however,
illustrate the fact that a small
amount of incoherent $c$-axis hopping cannot be completely
ruled out from consideration of the infrared data and that
this data can be understood quite well within Eliashberg
theory.
The last two curves (solid gray and dotted gray) are based on
BCS $d$-wave theory and are reproduced here to illustrate the
fact that such a theory is unable to explain the $c$-axis
data. The solid gray curve is with $t_\perp({\bf k}) =
t_\perp\cos^2(2\phi)$ and the dotted gray one for incoherent
$c$-axis transport. Compared with our Eliashberg results the
agreement with the data is poor.

\section{Conclusion}

We have considered the $c$-axis charge response as revealed
in the real part of the frequency dependent optical
conductivity $\sigma_{1c}(\omega)$. Results for a pure
BCS model, which includes only elastic impurity scattering,
show no agreement with the data in- or out-of-plane. A
generalized Eliashberg approach based on a spin fluctuation
mechanism leads to much better agreement.

In this approach the coherent delta function like part of the electron
spectral density which is sharply peaked at the quasiparticle
energy leads to a Drude type response in the
normal state and a BCS type response in the superconducting state.
The dominant incoherent background of the electron
spectral density provided by the coupling to the spin fluctuations,
however, leads to an additional boson assisted
region for the conductivity which is not strongly changed by
the onset of
superconductivity except that it is a shift by roughly $\Delta_0$.
This part of the conductivity, not included in BCS,
 contains the largest part of the optical
spectral weight (of order 75\%) and can be used to get information on
the underlying inelastic processes that presumably cause
superconductivity. The in-plane
conductivity reveals coupling to the $41\,$meV resonant peak
seen in spin polarized neutron scattering as well as to the
spin fluctuation background extending to high energies. For
coherent $c$-axis hopping the boson region 
in $\sigma_{1c}(\omega)$ has the same
form as it does in the in-plane case and this fact is 
largely born out
in the experimental data. This would not be the case if the
$c$-axis coupling were incoherent. Assuming pure incoherent
coupling gives no agreement
with the data although a best fit is
obtained with a small subdominant incoherent part in addition to a
dominant coherent part. An important conclusion coming out of
our analysis is that the $c$-axis conductivity data gives
independent confirmation for the form of the charge carrier
boson spectral density obtained solely from in-plane infrared
optical data.

While the infrared conductivity at higher energies is
dominated by the inelastic processes and not much affected
by a modest amount of elastic impurity scattering, the low
energy part is much more sensitive to impurities and depends
less directly on inelastic scattering. In this sense,
this frequency region can be partly understood within BCS theory
at least at low temperatures. One needs to remember, however,
that the spectral weight under the Drude-like curve is only
a fraction of the entire spectral weight and comes only
from the delta function-like part of the electron spectral density.
Also, even in this energy region, there are other
profound modifications
introduced by the inelastic scattering off the spin
fluctuation spectrum. For example all traces of the small
structures expected in $\sigma_1(\omega)$ at the gap and twice the
gap values are smeared out because, at such higher frequencies,
the inelastic scattering rate is already large.
Structures do appear in the data, however, at the energy of the gap
and of twice the gap plus once or twice the energy of the
spin resonance peak which is seen by inelastic neutron
scattering at $41\,$meV in optimally doped YBCO.

Finally, we point out that the spectral density for the
excitation spectrum that causes the superconductivity in
our approach which consists of a peak at $41\,$meV plus
a long nearly constant background extending to very large
energies while consistent with NAFFL of Pines and
coworkers,\cite{millis,mont} could also arise in other
microscopic mechanisms. Recently, the marginal Fermi liquid
model, proposed early on,\cite{abrah1,varma} in the development
of our understanding of the cuprates, and found to
reproduce very well many of the observed anomalous normal state
properties, has received new attention because it also fits
well the ARPES data. The excitation spectrum of the MFL is
quite similar to the one of NAFFL and has many of the features 
of our empirically determined spectrum provided its normal
state version is modified by the $41\,$meV resonance on
entering the superconducting state. In that sense the MFL is equally
consistent with optical data. Both, MFL and NAFFL, are
phenomenological models having similar excitation spectra
associated with the pairing and so cannot easily be distinguished
from optics.

\section*{Acknowledgment}

Research supported by the Natural Sciences and Engineering
Research Counsel of Canada (NSERC) and by the Canadian
Institute for Advanced Research (CIAR). We thank Dr. C.C. Homes
for discussions and for making his data available to us.

\newpage
\begin{figure}
\caption{Comparison of the real part of the in-plane
conductivity $\sigma_1(\omega)$ of a $d$-wave superconductor
(dashed line) and an $s$-wave superconductor (solid line).
Each have a gap of $\Delta_0 = 24\,$meV. For the $s$-wave
there is no absorption till $2\Delta_0$ while for the
$d$-wave with $\Delta_{\bf k} = \Delta_0\cos(2\phi)$ with
$\phi$ an angle in the two dimensional Brillouin zone, there
is a reduction as compared to the normal state Drude
(dotted curve). The slope of $\sigma_1(\omega)$ has a
maximum at $\Delta_0$ and $\sigma_1(\omega)$ has an additional small
structure at $2\Delta_0$.}
\label{f1}
\end{figure}
\begin{figure}
\caption{Comparison of the real part of the in-plane conductivity
$\sigma_1(\omega)$ vs. $\omega$ for various models. The grayed
solid line with a peak before $50\,$meV is BCS. The dash-dotted
line is an Eliashberg calculation
with an MMP spectral density peaked at $\omega_{\rm SF} = 20\,$meV.
The dashed line is the same but with $I^2\chi(\omega)$ used
instead of the MMP model. As described in the text this electron-%
boson spectral density $I^2\chi(\omega)$ has been determined through
a consideration
of the in-plane optical data. The dotted (Born) and dash-double-dotted
(unitary scattering) curves
include impurities in addition to the $I^2\chi(\omega)$ model for
inelastic scattering. The solid line
is the data of Homes {\it et al.}\cite{homes}}
\label{f2}
\end{figure}
\begin{figure}
\caption{The imaginary part of the conductivity $\omega\sigma_2(\omega)$
vs. $\omega$ for the various models described in Fig.~\ref{f2}. The
solid curve is the data. The dash-dotted curve is the result of
an Eliashberg calculation with an MMP model while the other curves
are based on the model $I^2\chi(\omega)$ which includes the
$41\,$meV resonance. These three curves are for the pure case
(only inelastic scattering, dashed line), and with some additional
elastic impurity scattering in Born (dotted) and unitary (dash-double-dotted)
limit with $t^+ = 0.32\,$meV and $\Gamma^+ = 0.63\,$meV respectively.}
\label{f2a}
\end{figure}
\begin{figure}
\caption{Second derivative $W(\omega)$ compared with input boson
spectral density $I^2\chi(\omega)$. The $41\,$meV peak in $I^2\chi(\omega)$
(gray squares) is clearly seen in $W(\omega)$ (solid line) as are the
tails at higher energies. In the energy region between 75 and $150\,$%
meV the van Hove singularities in the electronic density of states
show up added on to $\omega_r$ and distort the correspondence
between $W(\omega)$ and $I^2\chi(\omega)$.  }
\label{f2b}
\end{figure}
\begin{figure}
\caption{Comparison of the real part of the conductivity
$\sigma_1(\omega)$ vs.
$\omega$ for the in-plane case (dotted curve) and $c$-axis. The
solid line is for coherent $c$-axis hopping with
$t(\phi) = t_\perp\cos^2(\phi)$ with $\phi$ an
angle in the two dimensional CuO$_2$ Brillouin zone. The
dash-dotted is for incoherent tunneling with
$\vert V_1/V_0\vert = 1$. All curves are in the BCS model.}
\label{f3}
\end{figure}
\begin{figure}
\caption{Comparison between (solid) in-plane and out
of plane (dashed) real part of the $c$-axis conductivity
$\sigma_{1c}(\omega)$ vs. $\omega$ in an Eliashberg model
with our model carrier-boson spectral
density $I^2\chi(\omega)$ which includes the $41\,$meV spin
resonance. The dotted curve is $\sigma_{1c}(\omega)$
for a BCS $d$-wave
model with the same gap value as in the Eliashberg
work and is included for comparison.}
\label{f4}
\end{figure}
\begin{figure}
\caption{Comparison between in-plane (dotted) and out
of plane (solid) for the real part of $\sigma_1(\omega)$
vs. $\omega$. The data is from Homes {\it et al.}.%
\cite{homes}}
\label{f5}
\end{figure}
\end{document}